\documentclass{article}

\PassOptionsToPackage{numbers, compress}{natbib}
\usepackage[final]{neurips_2021_ml4ps}

\usepackage[utf8]{inputenc} % allow utf-8 input
\usepackage[T1]{fontenc}    % use 8-bit T1 fonts
\usepackage{hyperref}       % hyperlinks
\usepackage{url}            % simple URL typesetting
\usepackage{booktabs}       % professional-quality tables
\usepackage{amsfonts}       % blackboard math symbols
\usepackage{nicefrac}       % compact symbols for 1/2, etc.
\usepackage{microtype}      % microtypography
\usepackage{color,xcolor}         % colors

\usepackage{times}
\usepackage{epsfig}
\usepackage{graphicx,wrapfig}
\usepackage{amsmath}
\usepackage{amssymb}
\usepackage{subfigure}
\usepackage{cases}
\usepackage{overpic}
\usepackage{caption}
\usepackage{multirow}
\usepackage{tabularx}
\usepackage{arydshln}
\usepackage{array}
\usepackage{pifont}
\newcommand{\cmark}{\ding{51}}%
\newcommand{\xmark}{\ding{55}}%
\usepackage{enumitem}
\hypersetup{nolinks=True}
\usepackage{flushend}
\def\bx{{\bf x}}
\def\by{{\bf y}}
\def\bz{{\bf z}}

\def\0{{\bf 0}}
\def\1{{\bf 1}}

\def\bA{{\bf A}}

\def\bH{{\bf H}}

\def\bM{{\bf M}}

\def\bS{{\bf S}}

\DeclareMathOperator{\prox}{Prox}

\def\ie{\emph{i.e.}}
\def\eg{\emph{e.g.}}

\def\bsrnet{RBSRICNN}

\title{\bsrnet: Raw Burst Super-Resolution through Iterative Convolutional Neural Network}

\author{
  Rao Muhammad Umer\\
  Department of Computer Science\\
  University of Udine\\
  \texttt{engr.raoumer943@gmail.com}\\
  \And
  Christian Micheloni\\
  Department of Computer Science\\
  University of Udine\\
  \texttt{christian.micheloni@uniud.it} \\
}

\begin{document}

\maketitle

\begin{abstract}
    Modern digital cameras and smartphones mostly rely on image signal processing (ISP) pipelines to produce realistic colored RGB images. However, compared to DSLR cameras, low-quality images are usually obtained in many portable mobile devices with compact camera sensors due to their physical limitations. The low-quality images have multiple degradations \ie, sub-pixel shift due to camera motion, mosaick patterns due to camera color filter array, low-resolution due to smaller camera sensors, and the rest information are corrupted by the noise. Such degradations limit the performance of current Single Image Super-resolution (SISR) methods in recovering high-resolution (HR) image details from a single low-resolution (LR) image. In this work, we propose a Raw Burst Super-Resolution Iterative Convolutional Neural Network (\bsrnet\footnote{Our code and trained models are publicly available at \url{https://github.com/RaoUmer/RBSRICNN}}) that follows the burst photography pipeline as a whole by a forward (physical) model. The proposed Burst SR scheme solves the problem with classical image regularization, convex optimization, and deep learning techniques, compared to existing black-box data-driven methods. The proposed network produces the final output by an iterative refinement of the intermediate SR estimates. We demonstrate the effectiveness of our proposed approach in quantitative and qualitative experiments that generalize robustly to real LR burst inputs with onl synthetic burst data available for training.
\end{abstract}

%%%%%%%%% BODY TEXT
\section{Introduction}
The Burst Super-resolution is the task of fusing several low-resolution (LR) frames to produce a single high-resolution (HR) image. It is a fundamental low-level vision and image processing problem with various practical applications in satellite imaging, deforestation, environment and climate change monitoring, mobile photography, video enhancement, and security and surveillance imaging as well. In the last decade, most of the photos have been taken using built-in smartphone cameras, where the resulting low-quality images are inevitable and undesirable due to their physical limitations. Since the mobile cameras are small and versatile due to their compact camera sensors, there are several key limitations~\cite{delbracio2021mobilephotograph} of the mobile phone cameras as compared to a DSLRs \ie small sensor size, limited aperture, noise (\ie photon shot and read noise) and limited dynamic range, limited depth of field due to fixed aperture, limited zoom and color sub-sampling. As a result, the image quality is not comparable with that of DSLR cameras. Therefore, the focus is shifted towards software solutions to mitigate cameras hardware limitations.

Mathematically, the Burst SR problem is described as the following forward image observation model for degradation process:
\begin{equation}
    \small
    \by_i = \bM \bH \bS_i (\Tilde{\bx}) + \eta_i, ~~~~~ i = 1,\ldots, B
    \label{eq:degradation_model}
\end{equation}
where, $\by_i$ is the \emph{i-th} observed image of the LR burst $B$ images, $\bM$ is a \emph{mosaicking operator} that corresponds to the CFA (Color Filter Array) of a camera (usually Bayer), $\bH$ is a \emph{down-sampling operator} (\ie bilinear, bicubic, etc.) that resizes an HR image $\Tilde{\bx}$ by a scaling factor $r$, $\bS_i$ is an \emph{affine transformation} of the coordinate system of the image $\Tilde{\bx}$ (\ie translation and rotation), and $\eta_i$ is an additive \emph{heteroskedastic Gaussian noise} related to photon shot and read noise. Recovery of the latent HR image $\Tilde{\bx}$ from the LR burst belongs to the broad spectrum of inverse problems. The operators $\bM$ and $\bH$ are typically ill-conditioned, \ie singular. Further, this coupled with the presence of the noise $\eta_i$ and the affine transformation $\bS_i$ perturbing the measurements leads to an highly ill-posed nature of an inverse problem, where a unique solution does not exist. Burst photography pipeline with multi-frame super-resolution (MFSR) is the most common way to deal with such a scenario that generates the HR image from a low-quality burst of raw sensor images.

\begin{wrapfigure}{R}{25em}
\centering
\vspace{-0.5cm}
\includegraphics[width=8cm]{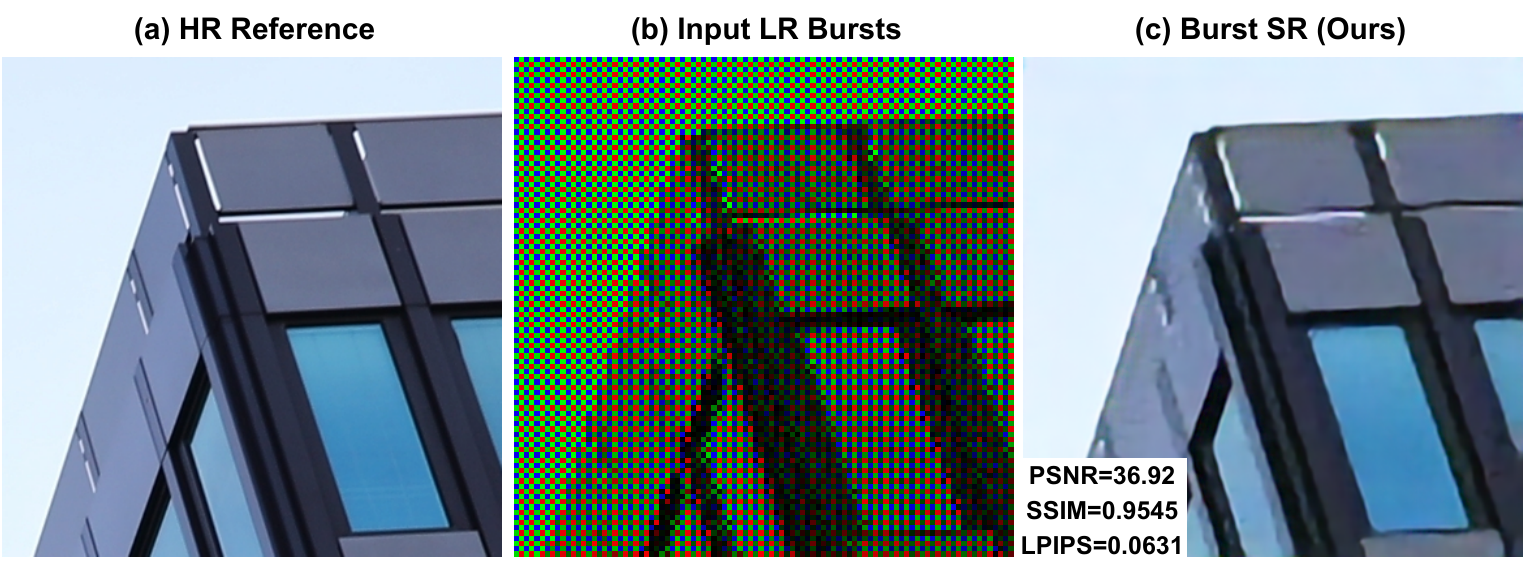}
\caption{An image from the Zurich RAW to RGB Dataset~\cite{ignatov2020rawtorgb} (testset), where we present (a) the ground truth HR reference image of size $384\times384\times3$, (b) the input LR bursts of size ($W\times H\times C\times B$) $48\times48\times4\times14$, and (c) the Burst SR output of size $384\times384\times3$ of our network (\bsrnet). All images are converted from raw sensor space to sRGB for visualization purpose.}
\label{fig:teaser}
\vspace{-0.3cm}
\end{wrapfigure}
Numerous works have been addressed towards the task of SISR~\cite{kim2016vdsrcvpr,Lim2017edsrcvprw,kai2017ircnncvpr,kai2018srmdcvpr,yuan2018unsupervised,Li2019srfbncvpr,zhang2019deep,srwdnet,Umer_2020_ICPR,luo2020unfoldingsr,zhou2020crossgraphsr,li2021laparsr,gou2020clearer,fan2020neuralir} and real-world SISR~\cite{ledig2017srgan,wang2018esrgan,lugmayr2019unsupervised,fritsche2019dsgan,umer2020srrescgan,umer2020srrescycgan,Umer2109srresstargan}. Due to the ill-posed nature of the SISR problem, the existing methods have limited performance to recover high frequency details through single image learned priors. On the other hand, the MFSR aims to recover the latent HR image using multiple LR frames by exploiting the additional signal information due to sub-pixel shifts, compared to the SISR methods. The recent works~\cite{gharbi2016deepjoint,kokkinos2019iterativecnn,kokkinos2019burstphotography,wronski2019handheldSR,deudon2020highresnet,bhat2021deepburstSR} have demonstrated the potential of MFSR methods that aim to fuse multiple LR frames to reconstruct a HR output. However, the deep learning based Raw Burst SR methods are black-box data-driven approaches with larger model size due to not directly model the image formation process, while our proposed scheme (\bsrnet) has the merit of interpretability and small model size with good reconstruction results by following the image observation model as shown in Fig.~\ref{fig:teaser}. Our proposed scheme exploits a powerful image regularization and large-scale optimization techniques by training a deep CNNs in an iterative manner to produce the final SR output with an iterative refinement of the intermediate SR estimates.  

%-------------------------------------------------------------------------
\section{Proposed Methodology}
\label{sec:proposed_method}
\subsection{Problem Formulation}
\label{sec:problem_formulation}
By referencing to the image observation model~\eqref{eq:degradation_model}, the recovery of $\bx$ from $\by_i$ mostly relies on the variational approach by combining the observations and prior knowledge, and the solution is obtained by minimizing the following objective function:
%\vspace{-0.3cm}
\begin{equation}
    \small
    \hat{\bx} = \underset{\mathbf{x}}{\arg \min }~\frac{1}{2\sigma^2B}\sum_{i=1}^{B}\|\by_i - \bM \bH \bS_i(\bx)\|_2^{2}+\lambda \mathcal{R}(\bx),
    \label{eq:eq1}
\end{equation}
where the first term corresponds to the data fidelity term that measures the closeness of the solution to the observations, while the second term (\ie $\mathcal{R}(\bx)$) corresponds to the regularization term that encodes any priors knowledge about the GT image, and $\lambda$ is the trade-off parameter that governs the compromise between the data fidelity and the regularizer term. The Eq.~\eqref{eq:eq1} can be also written in the following form:
\vspace{-0.3cm}
\begin{equation}
    \small
    \mathbf{J}(\bx) = \underset{\mathbf{x}}{\arg \min }~\frac{1}{2\sigma^2B}\|\by - \bA \bx \|_2^{2}+\lambda \mathcal{R}(\bx),
    \label{eq:eq2}
\end{equation}
where the ${\scriptstyle \bA = \bM \bH \bS }$ corresponds to the camera response. In the next section~\ref{subsec:obj_func_min}, we employ a proper optimization strategy to find the solution that minimizes the objective function~\eqref{eq:eq2} to get the required latent HR image.
\begin{figure}[!t]
\centering
\includegraphics[scale=0.85]{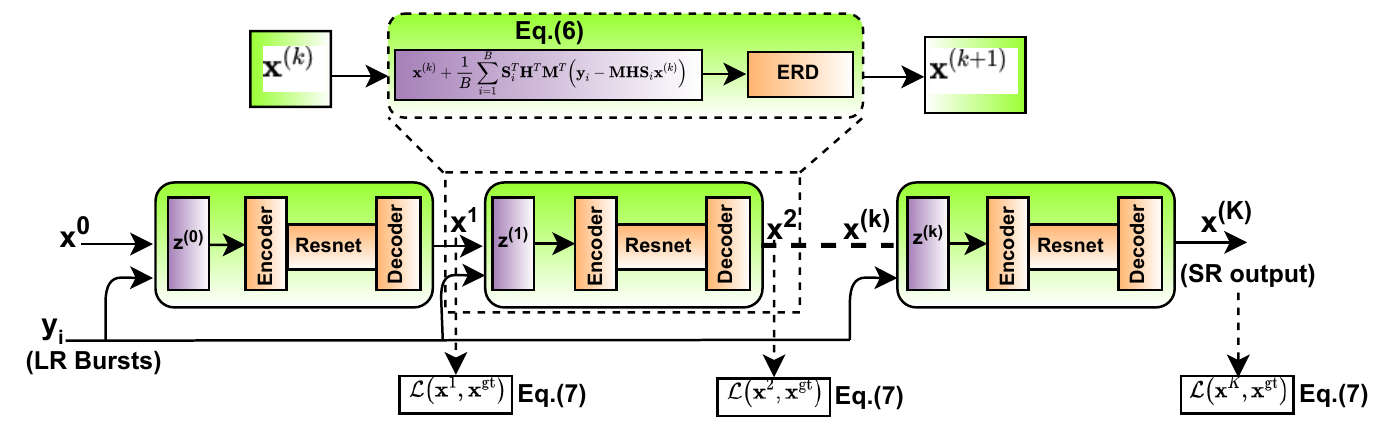}
\caption{Visualizes the structure of the our proposed iterative Raw Burst SR scheme. Given an LR burst images ($\by_i$), each network's stage produces a new estimate $\bx^{(k+1)}$ from the previous step estimate $\bx^{(k)}$. A single optimizer is used for all network stages with shared structures and parameters.}
\label{fig:bsricnn}
\vspace{-0.5cm}
\end{figure}
%\begin{wrapfigure}{R}{25em}
%    \includegraphics[width=\linewidth]{figs/bsricnn.pdf}
%    \caption{Visualizes the structure of the our proposed iterative Burst SR scheme. Given an LR burst images ($\by_i$), each network's stage produces a new estimate $\bx^{(k+1)}$ from the previous step estimate $\bx^{(k)}$. A single optimizer is used for all network stages with shared structures and parameters.}
%    \label{fig:bsricnn}
%    \vspace{-0.3cm}
%\end{wrapfigure}
\subsection{Optimization Scheme}
\label{subsec:obj_func_min}
In the Majorization-Minimization (MM)~\cite{hunter2004tutorial,figueiredo2007majorization,lefkimmiatis2011hessian} framework, an iterative algorithm for solving the minimization problem
$\hat{\bx} = \underset{\mathbf{x}}{\arg \min }~\mathbf{J}(\bx) ~\text{takes~the~form}~ \bx^{(k+1)} = \underset{\mathbf{x}}{\arg \min }~ \mathbf{Q}(\bx;{\bx}^{(k)}),$
where $\mathbf{Q}(\bx;{\bx}^{(k)})$ is the majorizer of the function $\mathbf{J}(\bx)$ at a fixed point ${\bx}^{(k)}$ by satisfying the following two conditions:
$\mathbf{Q}(\bx;{\bx}^{(k)})>\mathbf{J}(\bx), ~\forall \bx \ne \bx^{(k)} ~\text{and}~ \mathbf{Q}(\bx^{(k)};\bx^{(k)}) = \mathbf{J}(\bx^{(k)}).$
We need to upper-bound the $\mathbf{J}(\bx)$ by a suitable majorizer $\mathbf{Q}(\bx;{\bx}^{(k)})$ in order to get the required solution. Instead of minimizing the actual objective function \eqref{eq:eq2} due to its complexity, we minimize the majorizer function $\mathbf{Q}(.)$ to produce the next solution estimate $\bx^{(k+1)}$. By satisfying the above two properties of the majorizer function, iteratively minimizing $\mathbf{Q}(.;{\bx}^{(k)})$ also decreases the actual objective function $\mathbf{J}(.)$ \cite{hunter2004tutorial}. We write a quadratic majorizer for the complete objective function~\eqref{eq:eq2} as the following form:  
\begin{equation}
    \small
    \tilde{d}(\mathbf{x} ; \mathbf{x}^{(k)}) = \frac{1}{2 \sigma^{2} B}\|\mathbf{y}-\mathbf{A} \mathbf{x}\|_{2}^{2}+g(\mathbf{x}, \mathbf{x}^{(k)}),
    \label{eq:eq6}
\end{equation}
To start an estimate $\bx^{(k)}$, we have:
\begin{equation}
    \small
    g(\mathbf{x}, \mathbf{x}^{(k)})=\frac{1}{2 \sigma^{2} B}\left(\mathbf{x}-\mathbf{x}^{(k)}\right)^{T}\left[\alpha \mathbf{I}-\mathbf{A}^{T} \mathbf{A}\right]\left(\mathbf{x}-\mathbf{x}^{(k)}\right),
    \label{eq:eq7}
\end{equation}
where $\tilde{d}(\cdot, \cdot)$ is a distance function between $\bx$ and $\bx^{(k)}$. To get a valid majorizer, we need to satisfy the above two conditions such as $g(\mathbf{x}, \mathbf{y})>0, \forall \mathbf{x} \neq \mathbf{y} \text { and } g(\mathbf{x}, \mathbf{x})=0$. This suggests that $\alpha \mathbf{I}-\mathbf{A}^{T} \mathbf{A}$ must be a positive definite matrix, which only holds if $\alpha > \|\mathbf{A}^{T} \mathbf{A} \|_2  \Rightarrow \alpha \geq B$. Finally, we proceed with the following formulation to iteratively minimize the quadratic majorizer function $\mathbf{Q}(.)$ as:
\begin{equation}
\scriptsize
%\small
%\tiny
\hat{\bx}^{(k)} = \underset{\mathbf{x}}{\arg \min }~\mathbf{Q}(\bx;\bx^{(k)})
  = \tilde{d}\left(\mathbf{x} ; \mathbf{x}^{(k)}\right) + \lambda \mathcal{R}(\bx) 
  = \frac{\alpha}{2\sigma^2B}\|\bx - \bz^k\|_2^{2}+\lambda \mathcal{R}(\bx) + const. 
  = \prox_{(\lambda/\alpha\sigma^2)\mathcal{R}(.)}(\bz^k),
\label{eq:eq8}
\end{equation}
where ${\scriptstyle\bz^k = \bx^k + \bA^T(\by - \bA\bx^k) \Rightarrow \bz^k = \mathbf{x}^{(k)}+\frac{1}{B} \sum_{i=1}^{B} \mathbf{S}_{i}^{T} \mathbf{H}^{T}\mathbf{M}^{T}\left(\mathbf{y}_{i}-\mathbf{M H S}_{i} \mathbf{x}^{(k)}\right) }$ (see Fig.~\ref{fig:bsricnn}), and the \emph{const.} does not depend on $\bx$ and thus it is irrelevant to the optimization task. The $\prox_{(.)}$ is the proximal operator and it is computed as in~\cite{Umer_2020_ICPR}. Since the Eq.~\eqref{eq:eq8} is treated as the objective function of a denoising problem where $\bz$ is the noisy observation, we employ a deep denoising neural network to get the required solution estimate $\hat{\bx}^{(k)}$ by unrolling the network into $K$ finite stages. Moreover, in the Eq.~\eqref{eq:eq8}, we decouple the degradation operator $\bA$ from the latent image $\bx$ and now we need to tackle it with a less complex problem \ie denoising. For the fast convergence and less computational cost, we adopt a strategy similar to~\cite{kokkinos2019iterativecnn,Umer_2020_ICPR,kokkinos2019burstphotography}, where the trainable extrapolation weights $\mathbf{w^{(k)}}$ are learnt directly from the training data instead of the fixed ones~\cite{li2015accelerated}.

\subsection{Network Architecture and Training}
The proposed Raw Burst SR scheme is shown in Fig.~\ref{fig:bsricnn}. We unroll the proposed \bsrnet~into $K$ stages, where each stage computes the refined estimate of the solution. $\by_i$ is an input Raw LR burst, $\bx^0$ is an initial estimate, and $\bx^K$ is the final estimated SR image. We adapt the similar Encoder-Resnet-Decoder architecture as done in ~\cite{Umer_2020_ICPR}. Due to the iterative nature of our Burst SR approach, the network parameters are updated by using Truncated Backpropagation Through Time (TBPTT) algorithm as done in \cite{kokkinos2019iterativecnn,Umer_2020_ICPR} to train our network, where the sequence is unrolled into a small number of $k$-steps out of total $K$ and then the back-propagation is performed on the small $k$-steps. During the training, we use the following function to minimize the $\ell_1$-Loss between the estimated latent SR image ($\bx^{(k)}$) and ground-truth (GT) ($\bx^{(gt)}$) after k-steps as:
\begin{equation}
    \small
    \mathcal{L} = \arg\underset{\Theta}{\min}~\mathcal{L}(\mathbf{\Theta}) = \frac{1}{2} \sum \limits _{i = 1}^{N} \|\bx_i^{k} - \bx_{i}^{gt}\|_1
    \label{eq:l1loss}
\end{equation}
where $N$ is the mini-batch size and $\mathbf{\Theta}$ are the trainable parameters of our network. 

\begin{table*}[t]
    \centering
    \caption{\textbf{Comparison with others.}
    We compare our method with the common evaluation metrics (PSNR/SSIM/LPIPS). The quantitative SR results ($\times4$ upscale) are shown over the synthetic and real Burst SR test sets. The arrows indicate if high $\uparrow$ or low $\downarrow$ values are desired.}
    \vspace{-0.3cm}
    \resizebox{1.0\textwidth}{!}{
    \begin{tabular}{l@{\hskip 2mm} ccc@{\hskip 2mm}ccc @{\hskip 3mm} c @{\hskip 3mm} c @{\hskip 2mm}c}
    \toprule
    \multirow{2}{*}{\bf Burst SR Method}  & \multirow{2}{*}{\begin{tabular}[c]{@{}c@{}}~\bf\#Params~ \\ \bf ~[M]~\end{tabular}} & \multirow{2}{*}{\bf\#Conv2d} & \multicolumn{3}{c}{\bf Synthetic data} & \multicolumn{3}{c}{\bf Real data} & \multirow{2}{*}{\begin{tabular}[c]{@{}c@{}}~\bf Fine-tuned~ \\ \bf on Real data\end{tabular}} \\ %
    \cmidrule(r){4-6} \cmidrule(r){7-9} 
    &  &  & {\bf PSNR}$\uparrow$ & {\bf SSIM}$\uparrow$ & {\bf LPIPS}$\downarrow$ & {\bf PSNR}$\uparrow$ & {\bf SSIM}$\uparrow$ & {\bf LPIPS}$\downarrow$ &  \\
    \midrule
    {DeepJoint~\cite{gharbi2016deepjoint}+RRDB~\cite{wang2018esrgan}} & 17.26 & 371 & 33.25 & 0.881 & 0.195 & 42.13  & 0.957 & 0.088 & \cmark  \\ 
    {DeepBurstSR}~\cite{bhat2021deepburstSR} & 5.25 & 48 & 34.48 & \textbf{0.905} & \textbf{0.118} & \textbf{45.17}  & \textbf{0.978}  & \textbf{0.037} & \cmark  \\
    {HighRes-net}~\cite{deudon2020highresnet} & 1.11 & 25 & 34.30 & 0.891 & 0.170 & 43.99   & 0.972 & 0.051 & \cmark  \\ 
    {\bsrnet~(ours)} & \textbf{0.38} & \textbf{12} & \textbf{37.62} & 0.895 & 0.166  & 41.40 & 0.952 & 0.101  &  \xmark  \\ 
    \bottomrule
    \end{tabular}
    \label{tab:comp_sota}
    }
    \vspace{-0.3cm}
\end{table*}

%-------------------------------------------------------------------------
\section{Experiments}
%\vspace{-0.3cm}
\subsection{Training data and evaluation metrics}
%\vspace{-0.3cm}
We used $46,839$ and $1204$ sRGB images from the Zurich RAW to RGB dataset~\cite{ignatov2020rawtorgb} for the training and the validation, respectively. We generate the synthetic RAW LR bursts with their corresponding HR using the data generation code in \cite{bhat2021deepburstSR}, where the sRGB image is first converted to the \emph{Raw (linear) sensor} space using an inverse camera pipeline~\cite{brooks2019unprocessing}, then the LR burst is generated by applying random translations and rotations, followed by bilinear downsampling, further mosaicked and corrupted by random noise. For the evaluation of our method on the real dataset, we used the BurstSR testset containing $639$ real-world LR bursts, provided in the Burst SR challenge~\cite{bhat2021ntirebsr}. In the real LR bursts, each burst sequence contains 14 RAW images captured using a handheld smartphone camera using identical camera settings (\eg, exposure, ISO) resulting in a small random offset between the images within the burst. We evaluated the trained model under PSNR, SSIM, and LPIPS~\cite{zhang2018unreasonable} metrics. The quantitative Burst SR results are evaluated on the raw \emph{linear sensor} space. 

\subsection{Training description}
\label{appendix:train_desc}
%\vspace{-0.3cm}
We  estimate the warp matrix (\ie $\bS_{i}$, refers to the section-\ref{sec:proposed_method}) to align the bursts to the base/reference frame using the Enhanced Correlation Coefficient (ECC)~\cite{evangelidis2008eccmatrix} method as do in ~\cite{kokkinos2019burstphotography} for all experiments. For those bursts whose are not aligned by the ECC method, we keep them in the training and testing data by making the assumption of the identity matrix.   
For the training phase, we set the input Raw burst LR patch sizes as $48\times48\times4$ with their corresponding Raw HR patch sizes as $384\times384\times3$ by the scaling factor $\times4$. We use the LR burst  size of 14. We train the network for 368k iterations with a batch size of 2 using Adam optimizer~\cite{Kingma2015AdamAM} with parameters $\beta_1 =0.9$, $\beta_2=0.999$, and $\epsilon=10^{-8}$ without weight decay to minimize the loss~\eqref{eq:l1loss}. The learning rate is set to $10^{-3}$ for all iterations. We unroll the proposed network into $K$ stages, where we set $K$ as 10.
We implemented our method with Pytorch 1.7.1. The experiments are performed under Windows 10 with i7-8700 CPU with 32GB RAM and on NVIDIA GeForce RTX-3090 GPU with 24GB memory. The average running times (image per second on GPU) are $0.3350$ and $0.8838$ over the synthetic and real testsets, respectively.

\begin{figure*}[!t]
    \centering
    \includegraphics[width=\textwidth]{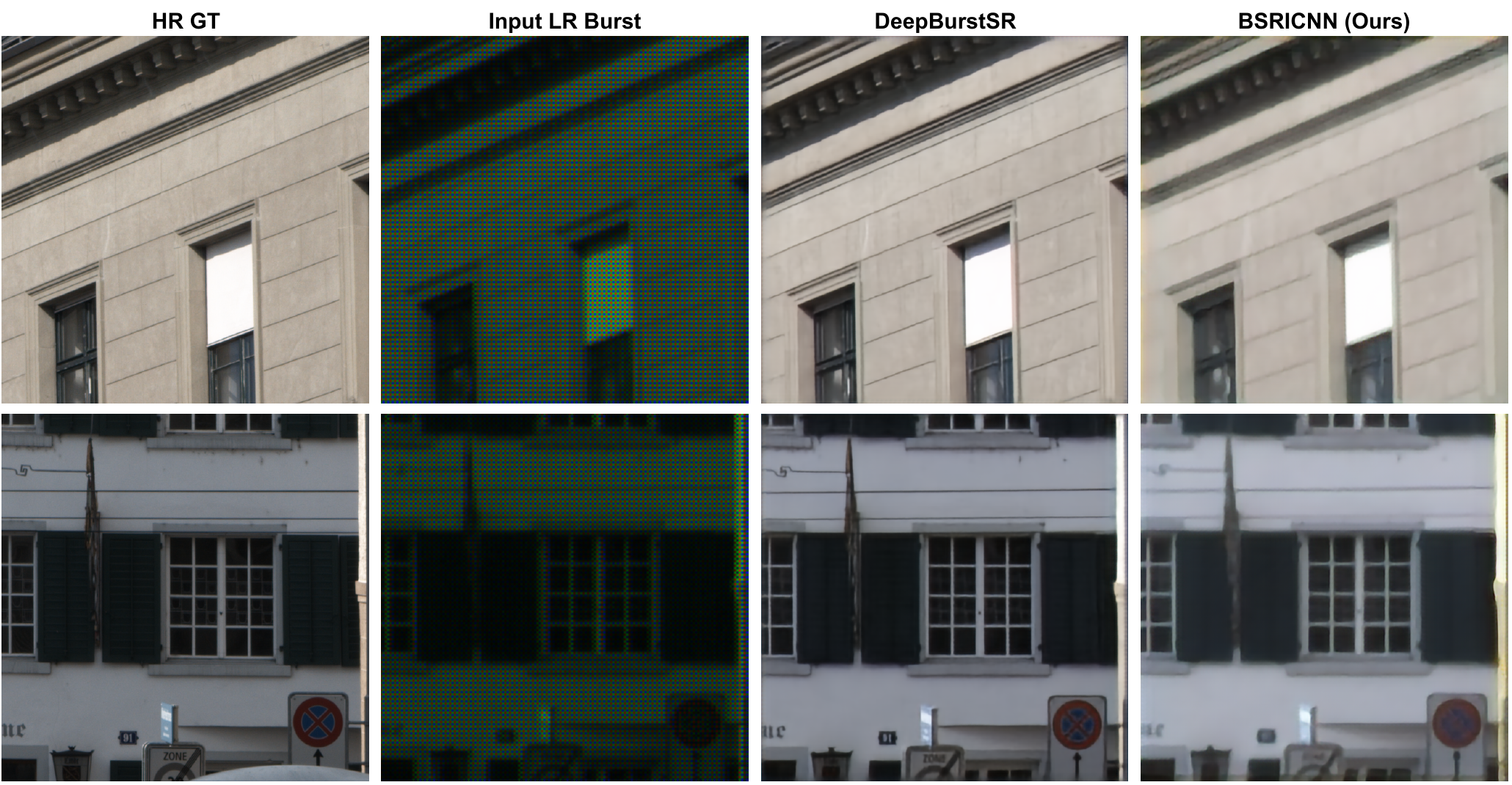}
    \caption{$\times4$ SR visual comparison of the proposed \bsrnet~method with the existing Burst SR methods on the real-world BurstSR~\cite{bhat2021deepburstSR} testset. All images are converted from the raw sensor space to sRGB for visualization purpose.}
    \label{fig:visual_comp}    
    \vspace{-0.5cm}
\end{figure*}

\subsection{Comparison with the Burst SR methods}
\label{sec:comp_sota}
We compared our method with existing Burst SR methods including DeepJoint~\cite{gharbi2016deepjoint}+RRDB~\cite{wang2018esrgan}, DeepBurstSR~\cite{bhat2021deepburstSR}, and HighRes-net ~\cite{deudon2020highresnet}. Table~\ref{tab:comp_sota} compares the quantitative SR results of our method with others over the synthetic and real Burst SR testsets. We have achieved good performance on the synthetic LR burst testset, while, the parameters and depth of our proposed network is much less than the other methods. On the real-world LR burst testset, the DeepBurstSR outperforms the others methods in terms of PSNR/SSIM/LPIPS, while we have a comparable performance, even though our method is not fine-tuned on the real training data with comparable visual results as shown in Fig.~\ref{fig:visual_comp}. The fine-tuning~\cite{bhat2021deepburstSR} increases the performance, but it also further requires significant additional labelled training data that is difficult to collect in practice, and it is also more computational expensive in terms of training time and hardware resources. We also participated in the NTIRE2021 Burst SR Challenges~\cite{bhat2021ntirebsr} and our method (\textbf{MLP\_BSR}) is ranked among other participants as well.

\begin{wraptable}{R}{5.0cm}
\vspace{-0.9cm}
  \caption{\textbf{Ablation study.} Impact of different number of input burst frames ($B$) and number of iterative steps ($K$). The quantitative results are reported on the synthetic burst testset.}
  \vspace{-0.3cm}
  \resizebox{0.37\textwidth}{!}{%
    \begin{tabular}{c @{\hskip 2mm} ccc @{\hskip 2mm} ccc}
    \toprule
    \multirow{2}{*}{\begin{tabular}[c]{@{}c@{}}~\bf Burst Size~ \\ \ \bf ($B$) \end{tabular}} & \multicolumn{3}{c}{\bf iterative steps ($K=5$)} & \multicolumn{3}{c}{ \bf iterative steps ($K=10$)} \\
    \cmidrule(r){2-4} \cmidrule(r){5-7}
     & \bf PSNR$\uparrow$ & \bf SSIM$\uparrow$ & \bf LPIPS$\downarrow$ & \bf PSNR$\uparrow$ & \bf SSIM$\uparrow$ & \bf LPIPS$\downarrow$ \\
    \midrule
    2  & 34.19 & 0.8790 & 0.2498 & 34.12 & 0.8777 & 0.2480 \\
    4  & 34.69 & 0.8852 & 0.2359 & 34.66 & 0.8842 & 0.2317 \\
    8  & 35.09 & 0.8887 & 0.2277 & 34.99 & 0.8876 & 0.2217 \\
    14 & 35.12 & 0.8896 & 0.2255 & 35.30 & 0.8903 & 0.2165 \\
    16 & \bf 35.21 & \bf 0.8907 & 0\bf .2232 & 35.30 & \bf0.8909 & 0.2168 \\
    32 & 35.23 & 0.8902 & 0.2236 & \bf35.41 & \bf0.8909 & \bf0.2159 \\
    \bottomrule
    \end{tabular}
    }
    \label{tab:ablation_table}
	\vspace{-0.7cm}
\end{wraptable}
%\vspace{-0.5cm}
\subsection{Ablation Study}
%\vspace{-0.3cm}
For our ablation study, we compared the impact of different numbers of input burst frames and iterative steps for the proposed Burst SR method in Table~\ref{tab:ablation_table}. It shows the generalization ability of our method to the larger input LR bursts and iterative steps, and effective utilization of the information from the multiple LR frames in order to improve burst SR performance. 

\subsection{Discussion and Limitation}
\label{appendix:disc_limit}
%\vspace{-0.3cm}
Our proposed method has a close connection to other proximal algorithms such as ISTA~\cite{daubechies2004ista} and FISTA~\cite{beck2009fista} that require the exact form of the employed regularizer such as Total Variation / Hessian Schatten-norm~\cite{lefkimmiatis2011hessian}. However, in our case, the regularizer is learned implicitly from the training data (\ie non-convex form) and therefore we do not have any assumptions regarding the explicit form of the regularizer. Our proposed algorithm acts as an inexact form of proximal gradient descent steps. Since we are estimating the warping matrix by using the ECC~\cite{evangelidis2008eccmatrix} method to align the observations to the reference frame, sometimes its estimation is imprecise that will introduce undesirable artifacts to the final SR result.
%-------------------------------------------------------------------------
%\vspace{-0.9cm}
\section{Conclusion}
%\vspace{-0.3cm}
We proposed a deep iterative Burst CNNs for a multi-frame super-resolution task by following the burst photography pipeline. The proposed burst SR scheme follows the burst physical model and solves the overall problem through iterative refinement of the intermediate solution estimates to get the final SR output. The proposed network exploits powerful image regularization, large-scale optimization, and deep learning techniques for multi-frame image restoration. Our model requires much less parameters and 2d convolution operation in comparison to other competing methods. Our method achieves good burst SR results in terms of the synthetic data as well as comparable visual quality of the real-world bursts with respect to the existing approaches.

%-------------------------------------------------------------------------
%\clearpage
%{\small
\bibliographystyle{ieee_fullname}
\bibliography{refs}
%}

\end{document}